\documentclass{fac}

\usepackage{hyperref}
\usepackage{graphicx}
\usepackage{amssymb}
\usepackage{amsfonts}
\usepackage{amsmath}
\usepackage{dsfont}
\usepackage{MnSymbol}
\usepackage{mathtools}
\usepackage{epsfig}
\usepackage{epstopdf}
\usepackage{lmerge}
\usepackage{tikz}
\usepackage{amsthm}
\ifprodtf \else \usepackage{latexsym}\fi

\usetikzlibrary{calc,decorations.pathmorphing,shapes,graphs}

\newcounter{sarrow}

\newcounter{sarrow1}
\newcommand\xnrsquigarrow[1]{%
\stepcounter{sarrow1}%
\mathrel{\begin{tikzpicture}[baseline= {( $ (current bounding box.south) + (0,-0.5ex) $ )}]
\node[inner sep=.5ex] (\thesarrow) {$\scriptstyle #1$};
\path[draw,<-,decorate,
  decoration={zigzag,amplitude=0.7pt,segment length=1.2mm,pre=lineto,pre length=4pt}]
    (\thesarrow1.south east) -- (\thesarrow1.south west);
    $\slashedarrowfill@\relbar\relbar/$
    \end{tikzpicture}}%
}

\makeatletter
\def\slashedarrowfill@#1#2#3#4#5{%
  $\m@th\thickmuskip0mu\medmuskip\thickmuskip\thinmuskip\thickmuskip
   \relax#5#1\mkern-7mu%
   \cleaders\hbox{$#5\mkern-2mu#2\mkern-2mu$}\hfill
   \mathclap{#3}\mathclap{#2}%
   \cleaders\hbox{$#5\mkern-2mu#2\mkern-2mu$}\hfill
   \mkern-7mu#4$%
}
\def\rightslashedarrowfillb@{%
  \slashedarrowfill@\relbar\relbar/\rightarrow}
\newcommand\xnrightarrow[2][]{%
  \ext@arrow 0055{\rightslashedarrowfillb@}{#1}{#2}}

\def\rightslashedarrowfille@{%
  \slashedarrowfill@\relbar\relbar/\twoheadrightarrow}
\newcommand\xntworightarrow[2][]{%
  \ext@arrow 0055{\rightslashedarrowfille@}{#1}{#2}}

\def\rightslashedarrowfillg@{%
  \slashedarrowfill@\relbar\relbar{\raisebox{.12em}{}}\twoheadrightarrow}
\newcommand\xtworightarrow[2][]{%
  \ext@arrow 0055{\rightslashedarrowfillg@}{#1}{#2}}

\def\rightslashedarrowfillx@{%
  \slashedarrowfill@\Relbar\Relbar/\rightrightarrows}
\newcommand\xnTworightarrow[2][]{%
  \ext@arrow 0055{\rightslashedarrowfillx@}{#1}{#2}}

\def\rightslashedarrowfilly@{%
  \slashedarrowfill@\Relbar\Relbar{\raisebox{.12em}{}}\rightrightarrows}
\newcommand\xTworightarrow[2][]{%
  \ext@arrow 0055{\rightslashedarrowfilly@}{#1}{#2}}

\pgfdeclareshape{slash underlined}
{
  \inheritsavedanchors[from=rectangle] 
  \inheritanchorborder[from=rectangle]
  \inheritanchor[from=rectangle]{north}
  \inheritanchor[from=rectangle]{north west}
  \inheritanchor[from=rectangle]{north east}
  \inheritanchor[from=rectangle]{center}
  \inheritanchor[from=rectangle]{west}
  \inheritanchor[from=rectangle]{east}
  \inheritanchor[from=rectangle]{mid}
  \inheritanchor[from=rectangle]{mid west}
  \inheritanchor[from=rectangle]{mid east}
  \inheritanchor[from=rectangle]{base}
  \inheritanchor[from=rectangle]{base west}
  \inheritanchor[from=rectangle]{base east}
  \inheritanchor[from=rectangle]{south}
  \inheritanchor[from=rectangle]{south west}
  \inheritanchor[from=rectangle]{south east}
  \inheritanchorborder[from=rectangle]
  \foregroundpath{
    \southwest \pgf@xa=\pgf@x \pgf@ya=\pgf@y
    \northeast \pgf@xb=\pgf@x \pgf@yb=\pgf@y
    \pgf@xc=\pgf@xa
    \advance\pgf@xc by .5\pgf@xb
    \pgf@yc=\pgf@ya
    \advance\pgf@xc by -1.3pt
    \advance\pgf@yc by -1.8pt
    \pgfpathmoveto{\pgfqpoint{\pgf@xc}{\pgf@yc}}
    \advance\pgf@xc by  2.6pt
    \advance\pgf@yc by  3.6pt
    \pgfpathlineto{\pgfqpoint{\pgf@xc}{\pgf@yc}}
    \pgfpathmoveto{\pgfqpoint{\pgf@xa}{\pgf@ya}}
    \pgfpathlineto{\pgfqpoint{\pgf@xb}{\pgf@ya}}
 }
}
\tikzset{nomorepostaction/.code=\let\tikz@postactions\pgfutil@empty}

\title[Draft of Process Algebra with Imperfect Actions]
      {Process Algebra with Imperfect Actions}

\author[Yong Wang]
    {Yong Wang\\
     Department of Computer Science and Technology,\\
     Faculty of Information Technology,\\
     Beijing University of Technology, Beijing, China\\
     }

\correspond{Yong Wang, Pingleyuan 100, Chaoyang District, Beijing, China.
            e-mail: wangy@bjut.edu.cn}

\pubyear{2023}
\pagerange{\pageref{firstpage}--\pageref{lastpage}}

\begin{document}
\label{firstpage}

\makecorrespond

\maketitle

\begin{abstract}
We discuss the deal of imperfectness of atomic actions in reality with the background of process algebras. And we show the applications of the imperfect actions in verification of computational systems.
\end{abstract}


\section{Introduction}

Process Algebras CCS \cite{CC} \cite{CCS} and ACP \cite{ACP} have bisimilarity-based interleaving semantics, while CTC and APTC \cite{APTC} are based on truly concurrent bisimilarities semantics and are generalizations of the corresponding CCS and ACP from interleaving semantics to truly concurrent semantics. 

In these process algebras, an atomic action $e$ from the set of atomic actions $\mathbb{E}$ is perfect, that is, $e$ can always execute itself and then terminate successfully, which is described by the following transition rule:

$$\frac{}{e\xrightarrow{e}\surd}$$

where $\xrightarrow{e}\surd$ is a predicate which means that it can execute $e$ and terminate successfully.

But, in reality, there is no any atomic action can be perfect and maybe execute with failures. An atomic action $e\in\mathbb{E}$ is always with failures for some probability $\pi$, and can be described by the following transition rule:

$$\frac{}{e\xrightarrow[\pi]{e}\surd}$$

Note that the above transition rule is not a standard rule which means that the execution of $e$ can be successful with a probability $\pi$ and failed with $1-\pi$. Now, the problem is how to introduce the intuitions of an imperfect action $e$ captured by the above transition rule into process algebras. 

Firstly, for each imperfect atomic action $e\in\mathbb{E}$, we introduce a corresponding perfect atomic action $\bar{e}$, that is, $\bar{e}$ can execute itself perfectly with a successful termination:

$$\frac{}{\bar{e}\xrightarrow{\bar{e}}\surd}$$

Secondly, the failures of execution of $e$ can be modelled by the deadlock $\delta$, which has no any behavior. 

Finally, the execution of $e$ can either be successful or be failed and should be equal to some choice between $\bar{e}$ and $\delta$. Can $e=\bar{e}+\delta$ be an axiom to be added into process algebras, where $+$ is the alternative composition operator?  Since $x+\delta=x$ is an axiom of process algebras, then we would have $e=\bar{e}+\delta=\bar{e}$. Exactly, this situation can be tackled by the probabilistic process algebras \cite{HTCPA}, i.e., there should be with the axiom $e=\bar{e}\boxplus_{\pi}\delta$ to be added into probabilistic process algebras, where $\boxplus_{\pi}$ is the probabilistic choice operator.

For the preliminaries on process algebras and probabilistic process algebras, we omit them and please refer to \cite{CC} \cite{CCS} \cite{ACP} \cite{APTC} \cite{HTCPA}. In the following sections, we take two verification examples of computational systems to show the usage of imperfect actions, based on probabilistic process algebra \cite{HTCPA}. Note that there are essential differences between the error handling of the system itself and the imperfect actions, the latter is unavoidable in real computational systems. For the simplicity, we do not introduce new kind of actions $\bar{e}$ for $e\in\mathbb{E}$, in the modelling of the computational systems, we just type $e\boxplus_{\pi}\delta$ instead of each occurrence of $e\in\mathbb{E}$, and assume that each $e\in\mathbb{E}$ is still perfect.

\section{Verification of Utopian Communication Protocol}

\begin{figure}
 \centering
 \includegraphics{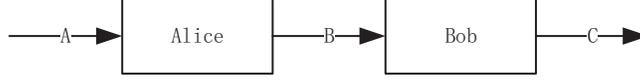}
 \caption{Utopian Communication Protocol}
 \label{ucp}
\end{figure}

Utopian Communication Protocol (UCP) is the almost simplest example shown in Fig. \ref{ucp}. Alice receives some data $d\in \Delta$ through the channel $A$, the corresponding reading action is denoted $r_A(d)$, then without any processing, she sends these data $d$ to Bob through the channel $B$, the corresponding sending action is denoted $s_B(d)$; Bob receives the data $d$ through the channel $B$ and the corresponding reading action is denoted $r_B(d)$, without any processing, he sending $d$ out though the channel $C$ and the corresponding sending action is denoted $s_C(d)$. Note that, all channels $A,B,C$ are perfect. Intuitively, it should be correct.

Alice's state transitions described by APTC are following.

$A=\sum_{d\in\Delta}(r_A(d)\boxplus_{\pi_1}\delta)\cdot A_1$

$A_1=\sum_{d\in\Delta}(s_B(d)\boxplus_{\pi_2}\delta)\cdot A_2$

$A_2=\circledS_{s_C(d)}\cdot A$

Bob's state transitions described by APTC are following.

$B=\circledS_{r_A(d)}\cdot B_1$

$B_1=\sum_{d\in\Delta}(r_B(d)\boxplus_{\pi_3}\delta)\cdot B_2$

$B_2=\sum_{d\in\Delta}(s_C(d)\boxplus_{\pi_4}\delta)\cdot B$

There is one communication function between Alice and Bob.

$$\gamma(r_B(d),s_B(d))\triangleq c_B(d)$$

Then the example can be represented by the following process term:

$$\tau_I(\partial_H(A\between B))$$

where $H=\{s_B(d),r_B(d)|d\in\Delta\}$, and $I=\{c_B(d)|d\in\Delta\}$.

\begin{eqnarray}
  A\between B&=&A\parallel B+A\mid B\nonumber\\
  &=&(\sum_{d\in\Delta}(r_A(d)\boxplus_{\pi_1}\delta)\cdot A_1)\parallel(\circledS_{r_A(d)}\cdot B_1)+(\sum_{d\in\Delta}(r_A(d)\boxplus_{\pi_1}\delta)\cdot A_1)\mid(\circledS_{r_A(d)}\cdot B_1)\nonumber\\
  &=&\sum_{d\in\Delta}r_A(d)\cdot (A_1\between B_1)\boxplus_{\pi_1}\delta\nonumber
\end{eqnarray}

\begin{eqnarray}
  \partial_H(A\between B)&=&\sum_{d\in\Delta}r_A(d)\cdot \partial_H(A_1\between B_1)\boxplus_{\pi_1}\delta\nonumber
\end{eqnarray}

\begin{eqnarray}
  A_1\between B_1&=&A_1\parallel B_1+A_1\mid B_1\nonumber\\
  &=&(\sum_{d\in\Delta}(s_B(d)\boxplus_{\pi_2}\delta)\cdot A_2)\parallel(\sum_{d\in\Delta}(r_B(d)\boxplus_{\pi_3}\delta)\cdot B_2)+(\sum_{d\in\Delta}(s_B(d)\boxplus_{\pi_2}\delta)\cdot A_2)\mid(\sum_{d\in\Delta}(r_B(d)\boxplus_{\pi_3}\delta)\cdot B_2)\nonumber\\
  &=&(\sum_{d\in\Delta}s_B(d))\parallel(\sum_{d\in\Delta}r_B(d))\cdot A_2\between B_2)\boxplus_{\pi_2}\delta\boxplus_{\pi_3}\delta+c_B(d)\cdot A_2\between B_2\boxplus_{\pi_2}\delta\boxplus_{\pi_3}\delta\nonumber
\end{eqnarray}

\begin{eqnarray}
  \partial_H(A_1\between B_1)&=&c_B(d)\cdot \partial_H(A_2\between B_2)\boxplus_{\pi_2}\delta\boxplus_{\pi_3}\delta\nonumber
\end{eqnarray}

\begin{eqnarray}
  A_2\between B_2&=&A_2\parallel B_2+A_2\mid B_2\nonumber\\
  &=&(\circledS_{s_C(d)}\cdot A)\parallel((s_C(d)\boxplus_{\pi_4}\delta)\cdot B)+(\circledS_{s_C(d)}\cdot A)\mid((s_C(d)\boxplus_{\pi_4}\delta)\cdot B)\nonumber\\
  &=&\sum_{d\in\Delta}s_C(d)\cdot (A\between B)\boxplus_{\pi_4}\delta\nonumber
\end{eqnarray}

\begin{eqnarray}
  \partial_H(A_2\between B_2)&=&\sum_{d\in\Delta}s_C(d)\cdot \partial_H(A\between B)\boxplus_{\pi_4}\delta\nonumber
\end{eqnarray}

By use of recursion, we can get the following linear recursion specification.

$X_1=\sum_{d\in\Delta}r_A(d)\cdot X_2\boxplus_{\pi_1}\delta$

$X_2=\sum_{d\in\Delta}c_B(d)\cdot X_3\boxplus_{\pi_2}\delta\boxplus_{\pi_3}\delta$

$X_3=s_C(d)\cdot X_1\boxplus_{\pi_4}\delta$

\begin{eqnarray}
  \tau_I(X_1)&=&\sum_{d\in\Delta}r_A(d)\cdot \tau_I(X_2)\boxplus_{\pi_1}\delta\nonumber\\
  &=&\sum_{d\in\Delta}r_A(d)\cdot \tau_I(X_3)\boxplus_{\pi_1}\delta\boxplus_{\pi_2}\delta\boxplus_{\pi_3}\delta\nonumber\\
  &=&\sum_{d\in\Delta}r_A(d)\cdot \sum_{d\in\Delta}s_C(d)\cdot \tau_I(X_1)\boxplus_{\pi_1}\delta\boxplus_{\pi_2}\delta\boxplus_{\pi_3}\delta\boxplus_{\pi_4}\delta)\nonumber
\end{eqnarray}

So, the Utopian Communication Protocol can exhibit desired external behaviors.

\section{Verification of Alternating Bit Protocol}

\begin{figure}
    \centering
    \includegraphics{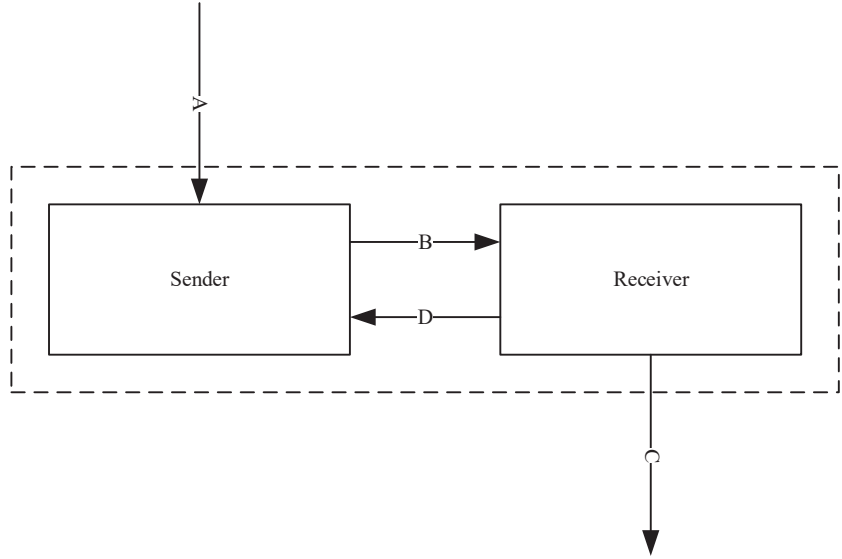}
    \caption{Alternating Bit Protocol}
    \label{abp}
\end{figure}

The Alternating Bit Protocol (ABP) protocol \cite{ABP} is used to ensure successful transmission of data through a corrupted channel. This success is based on the assumption that data can be resent an unlimited number of times, which is illustrated in Fig. \ref{abp}.

\begin{enumerate}
  \item Data elements $d_1,d_2,d_3,\cdots$ from a finite set $\Delta$ are communicated between a Sender and a Receiver.
  \item If the Sender reads a datum from channel $A$.
  \item The Sender processes the data in $\Delta$, formes new data, and sends them to the Receiver through channel $B$.
  \item And the Receiver sends the datum into channel $C$.
  \item If channel $B$ is corrupted, the message communicated through $B$ can be turn into an error message $\bot$.
  \item Every time the Receiver receives a message via channel $B$, it sends an acknowledgement to the Sender via channel $D$, which is also corrupted.
\end{enumerate}

The Sender attaches a bit 0 to data elements $d_{2k-1}$ and a bit 1 to data elements $d_{2k}$, when they are sent into channel $B$. When the Receiver reads a datum, it sends back the attached bit via channel $D$. If the Receiver receives a corrupted message, then it sends back the previous acknowledgement to the Sender.

Then the state transition of the Sender can be described by $APTC$ as follows.

\begin{eqnarray}
&&S_b=\sum_{d\in\Delta}(r_{A}(d)\boxplus_{\pi_1}\delta)\cdot T_{db}\nonumber\\
&&T_{db}=(\sum_{d'\in\Delta}((s_B(d',b)\boxplus_{\pi_2}\delta)\cdot \circledS^{s_{C}(d')})+(s_B(\bot)\boxplus_{\pi_3}\delta))\cdot U_{db}\nonumber\\
&&U_{db}=(r_D(b)\boxplus_{\pi_4}\delta)\cdot S_{1-b}+((r_D(1-b)\boxplus_{\pi_5}\delta)+(r_D(\bot)\boxplus_{\pi_6}\delta))\cdot T_{db}\nonumber
\end{eqnarray}

where $s_B$ denotes sending data through channel $B$, $r_D$ denotes receiving data through channel $D$, similarly, $r_{A}$ means receiving data via channel $A$, $\circledS^{s_{C}(d')}$ denotes the shadow of $s_{C}(d')$.

And the state transition of the Receiver can be described by $APTC$ as follows.

\begin{eqnarray}
&&R_b=\sum_{d\in\Delta}\circledS^{r_{A}(d)}\cdot R_b'\nonumber\\
&&R_b'=\sum_{d'\in\Delta}\{(r_B(d',b)\boxplus_{\pi_7}\delta)\cdot (s_{C}(d')\boxplus_{\pi_8}\delta)\cdot Q_b+(r_B(d',1-b)\boxplus_{\pi_9}\delta)\cdot Q_{1-b}\}+(r_B(\bot)\boxplus_{\pi_{10}}\delta)\cdot Q_{1-b}\nonumber\\
&&Q_b=((s_D(b)\boxplus_{\pi_{11}}\delta)+(s_D(\bot)\boxplus_{\pi_{12}}\delta))\cdot R_{1-b}\nonumber
\end{eqnarray}

where $\circledS^{r_{A}(d)}$ denotes the shadow of $r_{A}(d)$, $r_B$ denotes receiving data via channel $B$, $s_{C}$ denotes sending data via channel $C$, $s_D$ denotes sending data via channel $D$, and $b\in\{0,1\}$.

The send action and receive action of the same data through the same channel can communicate each other, otherwise, a deadlock $\delta$ will be caused. We define the following communication functions.

\begin{eqnarray}
&&\gamma(s_B(d',b),r_B(d',b))\triangleq c_B(d',b)\nonumber\\
&&\gamma(s_B(\bot),r_B(\bot))\triangleq c_B(\bot)\nonumber\\
&&\gamma(s_D(b),r_D(b))\triangleq c_D(b)\nonumber\\
&&\gamma(s_D(\bot),r_D(\bot))\triangleq c_D(\bot)\nonumber
\end{eqnarray}

Let $R_0$ and $S_0$ be in parallel, then the system $R_0S_0$ can be represented by the following process term.

$$\tau_I(\partial_H(\Theta(R_0\between S_0)))=\tau_I(\partial_H(R_0\between S_0))$$

where $H=\{s_B(d',b),r_B(d',b),s_D(b),r_D(b)|d'\in\Delta,b\in\{0,1\}\}\\
\{s_B(\bot),r_B(\bot),s_D(\bot),r_D(\bot)\}$

$I=\{c_B(d',b),c_D(b)|d'\in\Delta,b\in\{0,1\}\}\cup\{c_B(\bot),c_D(\bot)\}$.

Similar to the proof of the UCP protocol in the above section, the ABP protocol $\tau_I(\partial_H(R_0\between S_0))$ can exhibit desired external behaviors.

\label{lastpage}

\end{document}